\documentstyle[draft]{mn}

\title[Chemical evolution of the Magellanic Clouds]
      {Chemical Evolution of the Magellanic Clouds: analytical models}
\author[B.E.J. Pagel and G. Tautvai\v{s}ien\.{e}]
       {B.E.J. Pagel$^{1}$ and G. Tautvai\v sien\. e$^{2}$\\
       $^{1}$NORDITA, Blegdamsvej 17, Dk-2100 Copenhagen \O , Denmark 
(e-mail: pagel@nordita.dk)\\
       $^{2}$Institute of Theoretical Physics and Astronomy, Go\v{s}tauto 
12, Vilnius 2600, Lithuania (e-mail: taut@itpa.lt)}
\date{Received ........; in original form.............}

\newcommand{\eq}{\begin{equation}}
\newcommand{\en}{\end{equation}}

\begin{document}

\maketitle
\begin{abstract}
We have extended our analytical chemical evolution modelling ideas for 
the Galaxy (Pagel \& Tautvai\v sien\. e 1995, 1997) to the Magellanic 
Clouds.  Unlike previous authors (Russell 
\& Dopita 1992; Tsujimoto {\em et al.} 1995; Pilyugin 1996), we assume 
neither a steepened IMF nor selective galactic winds, since among the 
$\alpha$-particle elements only oxygen shows a large deficit relative 
to iron and a similar deficit is also found in Galactic supergiants. 
Thus we assume yields and time delays identical 
to those that we previously assumed for the solar neighbourhood. We 
include inflow and  non-selective galactic winds and consider both smooth
and bursting star formation rates, the latter giving a better fit to 
the age-metallicity relations. We predict essentially solar abundance 
ratios for primary elements and these seem to fit most of the data 
within their substantial scatter. Our LMC model also gives a remarkably 
good fit to the anomalous Galactic halo stars discovered by Nissen \& 
Schuster (1997). 

Our models predict current ratios of SNIa to core-collapse supernova 
rates enhanced by 50 per cent and 25 per cent respectively relative 
to the solar neighbourhood, in fair agreement with ratios found by 
Cappellaro {\em et al.} (1993) for Sdm-Im relative to Sbc galaxies, 
but these ratios 
are sensitive to detailed assumptions about the bursts and a still 
higher enhancement in the LMC has been deduced from X-ray studies of 
remnants by Hughes {\em et al.} (1995). The corresponding ratios 
integrated over time 
up to the present are slightly below 1, but they exceed 1 if one compares 
the Clouds with the Galaxy at times when it had the corresponding 
metallicities.

 \end{abstract}
\section{Introduction}

The histories of star formation and chemical evolution in the 
Magellanic Clouds exhibit 
distinct features in comparison to those of the disk of the Milky 
Way.  In both Clouds, as reviewed by Olszewski, Suntzeff \& Mateo (1996)
and Westerlund (1997), it seems that the majority of the stars are 
under 4 Gyr old, with a sprinkling of older stars.  In the LMC there is 
a sudden rise in the star formation rate (SFR) 2 to 4 Gyr ago (Elson, 
Gilmore \& Santiago 1997; Geisler {\em et al.} 1997) preceded by 
either a constant lower SFR (Geha 
{\em et al.} 1997) or possibly a virtual gap as manifested by the 
cluster age distribution (Da Costa 1991; van den Bergh 1991), 
itself preceded  
by the formation of a minority older population resembling the 
Galactic globular clusters although having disk-like kinematics, 
while in the SMC the SFR appears to have 
been more uniform (van den Bergh 1991; Olszewski, Suntzeff 
\& Mateo 1996). In the LMC, the
sudden rise in the SFR a few Gyr ago is reflected in a corresponding 
sudden rise in the metallicity, both from [Fe/H] in the clusters 
(Olszewski {\em et al.}; Geisler {\em et al.}) and from 
$\alpha$-particle elements in planetary 
nebulae (Dopita {\em et al.} 1997), while in the SMC there are fewer 
data, but some signs of a break in the 
age-metallicity relation (AMR) at an age between 2 and 4 Gyr (see below).  
From Olszewski {\em et al.}, the relative 
numbers of known clusters older and younger than 3 Gyr are 
15:100 in the   
LMC and 5:37 in the SMC, i.e. about equal proportions, but only one 
cluster is known in the LMC with age between 3 and 10 Gyr (and that one 
could be an interloper), while in the SMC there are 3 clusters known 
in the same interval. 

Many models have been put forward for the chemical evolution of the 
Magellanic Clouds, paying attention to the distinct Fe/O and Fe/$\alpha$
ratios which are generally found to be higher than in Galactic stars 
with the same metallicity, i.e. Fe/H. Gilmore \& Wyse (1991) pointed 
out that one way to get this effect is to assume separated star 
formation 
bursts, with SNIa contributing extra iron during quiescent intervals. 
Russell \& Dopita (1992) put forward a very detailed and comprehensive 
pair of models covering several element:element ratios previously 
determined by themselves and others and normalised to the solar 
neighbourhood (ISM and F supergiants) rather than the Sun.  This 
model assumed a smooth evolution in which the Clouds are built up by 
inflow of unprocessed gas according to certain time scales and a 
slightly steeper IMF than for the solar neighbourhood, which helps 
to bring down O/Fe and $\alpha$/Fe.  They noted that the observations 
seem to indicate [O/Fe] slightly lower than [$\alpha$/Fe], which might 
be a sign of O/$\alpha$ stellar yield ratios increasing with mass. 
They found but minor variations in heavy-element:iron ratios relative 
to the solar neighbourhood in the LMC, with the exception of an excess 
of Nd and  Sm which have a significant contribution from the r-process. 
In the SMC they found a still larger excess of  
Nd and the essentially pure  r-process elements Sm and Eu, averaging 
0.6 dex, and concluded that the main s-process had been less effective, 
and the r-process more effective, than in the solar neighbourhood. 
Russell \& Dopita's work referred to young objects (supergiants and the 
interstellar medium) and therefore did not include discussion of the 
age-metallicity relation. 

Another similarly comprehensive set of models was put forward by 
Tsujimoto {\em et al.} (1995), with the express purpose of explaining 
a relatively high SNIa/SNII ratio, deduced from the Fe/O ratio and 
supported by some direct counts of remnants, as had been proposed 
earlier by 
Barbuy, de Freitas Pacheco \& Castro (1994).  Tsujimoto {\em et al.} 
considered both smooth and bursting models.  From an optimal fit to 
14 elemental abundances from O up to Ni and smooth (bursting) chemical 
evolution 
models they deduced values of 0.15, 0.24 (0.21) and 0.3 (0.28) for the 
ratio, 
integrated over time, of SNIa/SN(II+Ib+Ic) in the solar neighbourhood, 
the LMC and the SMC respectively, and suggested that they 
originate from Salpeter-like power-law exponents in the IMF of 
respectively 1.33, 1.71 (1.62) and 1.88 (1.84).  These steeper IMFs
(between fixed upper and lower mass limits), also postulated by 
Russell \& Dopita, 
then also  serve to produce the subsolar metallicities in the Magellanic 
Clouds at the corresponding gas fractions and give reasonable 
age-metallicity relations, especially with the burst models. 

The hypothesis of a slightly steeper IMF compared to the Galaxy 
is neither supported nor ruled out by direct star counts 
(Hill, Madore \& Freedman 1994 and references therein). 
An alternative way to explain low metallicities is to assume outflow
(e.g. Carigi {\em et al.} 1995).  Such outflow can be assumed to be 
either homogeneous or selective, the latter case being associated 
with starbursts and leading to 
enhancement of the Fe/O ratio, among others (e.g. Marconi, Matteucci 
\& Tosi 1994). Such a model has been put forward for the LMC by 
Pilyugin (1996), who points out that a closed bursting model, like 
those of Gilmore \& Wyse, will produce the same final oxygen abundance 
at a given gas fraction as a smooth model, so that some other factor 
is needed to provide the reduction in metallicity compared to the 
solar neighbourhood with the same gas fraction.  Given the absence 
of compelling evidence for a change in the IMF, 
Pilyugin postulates selective winds 
associated with star formation bursts; an initial square-wave burst 
in the first 0.2 Gyr, accounting for 8 per cent of the stars, is followed 
by a hiatus until 9 Gyr, when a major burst or series of bursts 
begins, ending after 12.5 Gyr (0.5 Gyr before present) and accounting 
for the remainder, in accordance with the age distribution of the 
LMC clusters. A model with both homogeneous and selective galactic 
winds (with a selective expulsion of 75 per cent of ejecta from massive 
stars and a homogeneous expulsion of a mass equal to that being made 
into stars) is found to give good fits to the AMR  and to [Fe/O].  

Unfortunately, the assessment of [Fe/O] and, to a lesser extent, 
other abundance ratios is dependent on which Galactic standards are 
used (cf Pagel 1992). While it is certainly true that [Fe/O] is not 
negative in the Clouds, it is not strongly positive either when 
compared to Galactic supergiants instead of the Sun, and because of 
various systematic effects the supergiants may provide a better 
standard.  Among $\alpha$-particle elements, notably Mg and Ca, and 
for Ti which seems to behave like the $\alpha$-elements among Galactic 
stars, the offset from either solar or local supergiant ratios is 
insignificantly different from zero, compared to small positive values
in the Milky Way for the same [Fe/H] (Edvardsson {\em et al.} 1993). 
There is thus some difference, but not necessarily large enough to 
demand either a changed IMF or selective galactic winds.  Some sort of 
winds are probably present, as witness the Magellanic stream, and 
these may well account for the relatively low metallicities, and we 
agree with Russell \& Dopita and with Tsujimoto {\em et al.} in 
assuming a significant effect of inflow, since this is needed, at 
least in smooth models with reasonable star formation laws, to 
account for the `G-dwarf' problem posed by the relatively small 
number of old objects.  The existence of separated star formation 
bursts is well attested in the LMC by population studies of both clusters
 and field stars, and is supported in both Clouds by the AMRs.  

With this background, it seems worth while to investigate how well 
the chemical evolution of the Magellanic Clouds can be modelled 
with yields and time-delays equal to those applying to the solar 
neighbourhood, but with inflow and a homogeneous wind.  In two 
previous papers (Pagel \& Tautvai\v{s}ien\.{e} 1995, 1997) we 
developed analytical models giving a fair fit to stellar abundances 
of primary elements in the solar neighbourhood, and in this paper 
we shall attempt to apply the same ideas to the Clouds. 

\section{The model}

\subsection{General description}
\input epsf 
\begin{figure}
\epsfxsize=\hsize
\epsfbox[85 85 510 297]{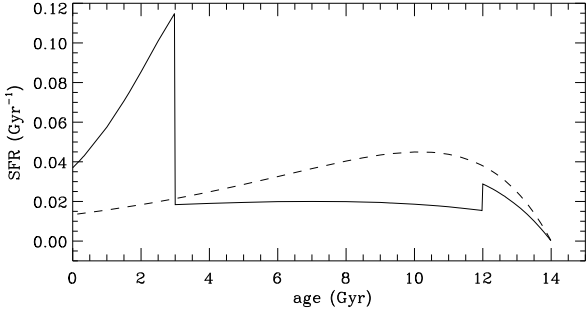} 
\caption{SFR history for the LMC according to our model.  The full
curve shows the bursting model according to Table 1, while the 
broken-line curve shows a smooth model with $u = 0.18t$.} 
\label{fig1} 
\end{figure} 

\begin{figure}
\epsfxsize=\hsize 
\epsfbox[85 85 510 297]{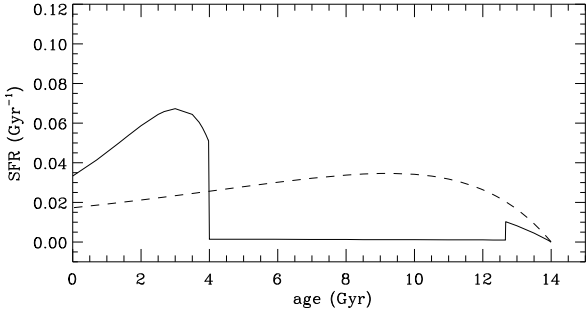}
\caption{SFR history for the SMC according to our model.  The full 
curve shows the bursting model according to Table 2, while the
broken-line curve shows a smooth model with $u=0.115t$.}
\label{fig2}
\end{figure} 

In agreement with Russell \& Dopita (1992) and Tsujimoto {\em et al.}
(1995), we assume the Clouds to have been built up by gradual infall
of unprocessed material.  This helps to alleviate the `G-dwarf'
problem that would otherwise arise from the small relative proportion of
old, metal-poor stars.  Our formalism is an adaptation of that of
 Pagel (1997) as detailed below;
 like Tsujimoto {\em et al.}, we assume linear laws of star formation
 and  investigate both smooth and bursting models. Like Pilyugin (1996),
we assume yields and time delays identical to those which apply to the 
solar neighbourhood and appeal to galactic winds to explain the low 
metallicities of the Clouds in relation to their current gas fractions; 
we take the specific numbers from our two 
previous papers.  Unlike Pilyugin, however, we  ignore selective winds 
and assume  just a non-selective wind proportional to the SFR, similar 
to the model of Hartwick (1976) for the Galactic halo. 

\subsection{The formalism} 

\begin{figure}
\epsfxsize=\hsize 
\epsfbox[85 85 510 297]{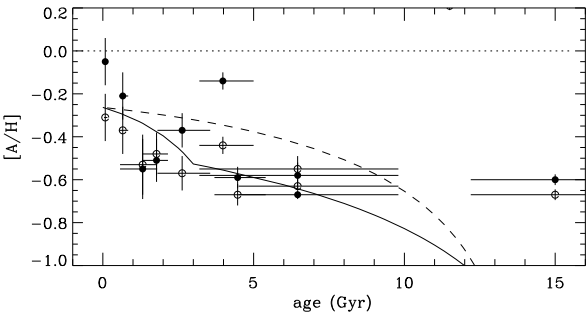} 
\caption{Age-abundance relation for $\alpha$-elements in the LMC. 
Curves are as in Fig. 1.  Data points are from Dopita 
{\em et al.} (1997): open circles represent oxygen while filled 
circles represent an average of Ne, S and Ar.} 
\label{fig3} 
\end{figure} 

We assume a linear or quasi-linear star formation law 
\eq
\frac{ds}{du}= g, 
\en
where $s$ is the mass in stars (+ remnants), $g$ is the mass of gas and 
$u=\int_{0}^{t}\omega(t')dt'$ where $t$ is time and $\omega$ is the 
inverse time scale for star formation, assumed 
constant in the `smooth' models.  In `bursting' models, $\omega$ is assumed 
to be constant over certain time periods, between which it changes
discontinuously. Inflow is assumed to occur at a rate  
\eq
f(t)=\omega(t)\, e^{-u},
\en 
and outflow at a rate 
\eq
e(t)=\eta\frac{ds}{dt}=\eta\,\omega g,
\en 
with $\eta =$ const., so that the gas mass and total  mass satisfy 
the differential equations 
\eq
\frac{dg}{du}=e^{-u}-(1+\eta)g 
\en 
and 
\eq
\frac{dm}{du}=e^{-u}-\eta g 
\en 
respectively.  From Eqs. (4), (5), the gas and total masses 
evolve according to
\eq 
g(u)=\left[e^{-u}-e^{-(1+\eta)u}\right]/\eta;
\en
\eq 
m(u)= \left[1-e^{-(1+\eta)u}\right]/(1+\eta);    
\en
the star mass is 
\eq
s(u)=m(u)-g(u)  
\en 
and the gas fraction is 
\eq
\mu(u) = g(u)/m(u).  
\en 
Using the instantaneous recycling approximation, the abundance of 
a promptly produced primary element such as O or Mg, 
in units of its yield (assumed constant), satisfies 
\eq
\frac{d}{du}(gz_{1})=g[1-(1+\eta)z_{1}], 
\en 
which has the solution
\eq
z_{1}(u)=\frac{1}{\eta}-\frac{u}{e^{\eta u}-1}. 
\en 
Results of Eqs. (1) to (11) for our assumed model para\-meters 
are given in the first seven columns of Tables 1 and 2. 
Figs 1 and 2 show the resulting 
SFR histories for both smooth and bursting models; the rough
coincidence in time between our assumed bursts in the two Clouds
suggests that they may have resulted from some mutual interaction.  
For the LMC, 
our assumed value of 1 for $\eta$ agrees with that adopted by 
Pilyugin and the current star formation rate is about average; 
this may be compared with the remark by Westerlund (1997) that 
the current SFRs in both Clouds deduced from H$\alpha$ emission
may indicate that their current SFRs are below their averages. 
Thus the decline that we assume in the last 3 Gyrs from the peak 
of the recent burst, resulting from our assumption of Eq. (1) 
with a constant $\omega$ in the relevant interval, 
seems to be fairly realistic.   The final gas fraction adopted 
for the SMC is in accordance with the number given by Westerlund;
for the LMC it is somewhat larger than his figure of $<8\%$, but 
in accordance with that adopted in other models in the literature
and similar to some estimates for the solar neighbourhood.
Fig 3 
shows the age-abundance relation in the LMC for O, Ne, S and Ar, 
assumed to be instantaneously produced with a yield of 0.7 
times their solar abundance, and confirms that a bursting model 
gives a better fit than a smooth one.  

\begin{table*}
\centering
\caption{Bursting model for the LMC; $\langle\omega\rangle = 0.18$ 
Gyr $^{-1}$; $\eta = 1$.}
\begin{tabular}{ccccccccccccc}
$\omega$ & $t$  & $u$ & $g$ & $m$ & $\mu$ & $z_{1}$& $z_{2}$& $z_{2}$& $z_{2}$& $z_{2}$ & [O/H] & [Fe/H]\\
Gyr$^{-1}$&Gyr&&&&&&0.023&0.037&1.33&2.67\\ 
&&&&&&\\
0.15 &  0.023  & 0.003 & 0.003 & 0.003 &   1.00 & 0.002 & 0.000 & 0.000 & 0.000 & 0.000 & --2.92 & --3.31  \\
      &  0.037 & 0.006 & 0.006 & 0.006 &   1.00 & 0.003 & 0.000 & 0.000 & 0.000 & 0.000 & --2.71 & --3.11  \\
      &  0.500 & 0.075 & 0.067 & 0.070 &   0.96 & 0.037 & 0.034 & 0.032 & 0.000 & 0.000 & --1.59 & --1.98  \\
      &  0.900 & 0.135 & 0.110 & 0.118 &   0.93 & 0.066 & 0.063 & 0.061 & 0.000 & 0.000 & --1.34 & --1.73  \\
      &  1.330 & 0.200 & 0.148 & 0.165 &   0.90 & 0.096 & 0.094 & 0.092 & 0.000 & 0.000 & --1.17 & --1.57  \\
 0.15 &  2.000 & 0.300 & 0.192 & 0.226 &   0.85 & 0.143 & 0.140 & 0.139 & 0.022 & 0.000 & --1.00 & --1.31  \\
&&&&&&\\
 0.08 &  2.023 & 0.302 & 0.193 & 0.227 &   0.85 & 0.143 & 0.143 & 0.141 & 0.024 & 0.000 & --1.00 & --1.30  \\
      &  2.037 & 0.303 & 0.193 & 0.227 &   0.85 & 0.144 & 0.143 & 0.143 & 0.025 & 0.000 & --1.00 & --1.30  \\
      &  2.670 & 0.354 & 0.209 & 0.254 &   0.83 & 0.166 & 0.166 & 0.165 & 0.073 & 0.000 & --0.93 & --1.11  \\
      &  3.330 & 0.406 & 0.222 & 0.278 &   0.80 & 0.190 & 0.189 & 0.188 & 0.135 & 0.007 & --0.88 & --0.96  \\
      &  4.670 & 0.514 & 0.240 & 0.321 &   0.75 & 0.235 & 0.234 & 0.234 & 0.184 & 0.046 & --0.78 & --0.84  \\
      &  6.000 & 0.620 & 0.249 & 0.355 &   0.70 & 0.278 & 0.278 & 0.277 & 0.234 & 0.116 & --0.71 & --0.75  \\
      &  7.000 & 0.700 & 0.250 & 0.377 &   0.66 & 0.310 & 0.309 & 0.309 & 0.270 & 0.167 & --0.66 & --0.70  \\
      &  8.000 & 0.780 & 0.248 & 0.395 &   0.63 & 0.340 & 0.339 & 0.339 & 0.306 & 0.215 & --0.62 & --0.65  \\
      &  9.000 & 0.860 & 0.244 & 0.411 &   0.59 & 0.369 & 0.369 & 0.369 & 0.341 & 0.261 & --0.59 & --0.61  \\
      & 10.000 & 0.940 & 0.238 & 0.424 &   0.56 & 0.397 & 0.397 & 0.397 & 0.375 & 0.306 & --0.56 & --0.57  \\
 0.08 & 11.000 & 1.020 & 0.231 & 0.435 &   0.53 & 0.425 & 0.425 & 0.425 & 0.408 & 0.349 & --0.53 & --0.54  \\
&&&&&&\\
 0.50 & 11.023 & 1.032 & 0.229 & 0.437 &   0.53 & 0.429 & 0.419 & 0.419 & 0.402 & 0.345 & --0.52 & --0.54  \\
      & 11.037 & 1.039 & 0.229 & 0.437 &   0.52 & 0.431 & 0.421 & 0.415 & 0.399 & 0.342 & --0.52 & --0.54  \\
      & 11.500 & 1.270 & 0.202 & 0.461 &   0.44 & 0.504 & 0.498 & 0.495 & 0.320 & 0.280 & --0.45 & --0.56  \\
      & 12.330 & 1.685 & 0.151 & 0.483 &   0.31 & 0.616 & 0.616 & 0.615 & 0.256 & 0.236 & --0.37 & --0.55  \\
      & 12.500 & 1.770 & 0.141 & 0.486 &   0.29 & 0.637 & 0.637 & 0.637 & 0.359 & 0.234 & --0.35 & --0.48  \\
      & 12.670 & 1.855 & 0.132 & 0.488 &   0.27 & 0.007 & 0.657 & 0.657 & 0.455 & 0.234 & --0.35 & --0.43  \\
      & 13.000 & 2.020 & 0.115 & 0.491 &   0.23 & 0.691 & 0.693 & 0.695 & 0.626 & 0.239 & --0.32 & --0.34  \\
      & 13.670 & 2.355 & 0.086 & 0.496 &   0.17 & 0.753 & 0.757 & 0.760 & 0.915 & 0.270 & --0.28 & --0.23  \\
      & 13.800 & 2.420 & 0.081 & 0.496 &   0.16 & 0.764 & 0.769 & 0.771 & 0.963 & 0.422 & --0.27 & --0.21  \\
 0.50 & 14.000 & 2.520 & 0.074 & 0.497 &   0.15 & 0.780 & 0.785 & 0.787 & 1.033 & 0.645 & --0.26 & --0.19  \\
\end{tabular}
\end{table*}

\begin{table*}
\centering
\caption{Bursting model for the SMC; $\langle\omega\rangle = 0.115$  
Gyr$^{-1}$; $\eta = 2$.}
\begin{tabular}{ccccccccccccc}
$\omega$ & $t$ & $u$ & $g$ & $m$ & $\mu$ & $z_{1}$& $z_{2}$& $z_{2}$& $z_{2}$& $z_{2}$ & [O/H] & [Fe/H]\\
Gyr$^{-1}$&Gyr&&&&&&0.023&0.037&1.33&2.67\\
&&&&&&\\
 0.10 &  0.023 & 0.002 & 0.002 & 0.002 &   1.00 & 0.001 & 0.000 & 0.000 & 0.000 & 0.000 & --3.10 &  --3.49 \\
      &  0.037 & 0.004 & 0.004 & 0.004 &   1.00 & 0.002 & 0.000 & 0.000 & 0.000 & 0.000 & --2.89 &  --3.29 \\
      &  0.500 & 0.050 & 0.045 & 0.046 &   0.97 & 0.025 & 0.023 & 0.021 & 0.000 & 0.000 & --1.76 &  --2.16 \\
      &  1.000 & 0.100 & 0.082 & 0.086 &   0.95 & 0.048 & 0.046 & 0.045 & 0.000 & 0.000 & --1.47 &  --1.87 \\
0.10  &  1.330 & 0.133 & 0.102 & 0.110 &   0.93 & 0.064 & 0.062 & 0.061 & 0.000 & 0.000 & --1.35 &  --1.75 \\
&&&&&&\\
0.01  &  1.353 & 0.133 & 0.102 & 0.110 &   0.93 & 0.064 & 0.064 & 0.063 & 0.000 & 0.000 & --1.35 &  --1.75 \\
      &  1.367 & 0.133 & 0.102 & 0.110 &   0.93 & 0.064 & 0.064 & 0.064 & 0.000 & 0.000 & --1.35 &  --1.75 \\
      &  2.660 & 0.146 & 0.110 & 0.118 &   0.93 & 0.070 & 0.070 & 0.070 & 0.067 & 0.000 & --1.31 &  --1.32 \\
      &  4.000 & 0.160 & 0.117 & 0.127 &   0.92 & 0.076 & 0.076 & 0.076 & 0.072 & 0.002 & --1.28 &  --1.29 \\
      &  6.000 & 0.180 & 0.126 & 0.139 &   0.91 & 0.085 & 0.085 & 0.085 & 0.081 & 0.019 & --1.23 &  --1.24 \\
      &  8.100 & 0.201 & 0.135 & 0.151 &   0.90 & 0.094 & 0.094 & 0.094 & 0.089 & 0.034 & --1.18 &  --1.20 \\
0.01  & 10.000 & 0.220 & 0.143 & 0.161 &   0.89 & 0.102 & 0.102 & 0.102 & 0.097 & 0.047 & --1.15 &   --1.16\\
&&&&&&\\
0.35  & 10.023 & 0.228 & 0.146 & 0.165 &   0.88 & 0.105 & 0.098 & 0.098 & 0.093 & 0.045 & --1.13 &  --1.16 \\
      & 10.037 & 0.233 & 0.147 & 0.168 &   0.88 & 0.107 & 0.100 & 0.095 & 0.091 & 0.044 & --1.12 &  --1.17 \\
      & 10.200 & 0.290 & 0.165 & 0.194 &   0.85 & 0.131 & 0.124 & 0.121 & 0.070 & 0.035 & --1.04 &  --1.18 \\
      & 10.500 & 0.395 & 0.184 & 0.231 &   0.80 & 0.172 & 0.167 & 0.164 & 0.048 & 0.024 & --0.92 &  --1.17 \\
      & 11.000 & 0.570 & 0.192 & 0.273 &   0.70 & 0.232 & 0.229 & 0.227 & 0.030 & 0.017 & --0.79 &  --1.11 \\
      & 11.330 & 0.685 & 0.188 & 0.291 &   0.65 & 0.267 & 0.265 & 0.263 & 0.024 & 0.014 & --0.73 &  --1.07 \\
      & 12.000 & 0.920 & 0.168 & 0.312 &   0.54 & 0.326 & 0.326 & 0.326 & 0.189 & 0.012 & --0.64 &  --0.77 \\
      & 12.670 & 1.154 & 0.142 & 0.323 &   0.44 & 0.373 & 0.373 & 0.374 & 0.337 & 0.012 & --0.58 &  --0.61 \\
      & 13.300 & 1.375 & 0.118 & 0.328 &   0.36 & 0.406 & 0.408 & 0.409 & 0.451 & 0.244 & --0.55 &  --0.52 \\
0.35  & 14.000 & 1.620 & 0.095 & 0.331 &   0.29 & 0.434 & 0.436 & 0.438 & 0.550 & 0.494 & --0.52 &  --0.45 \\
\end{tabular}
\end{table*}

For elements (or components thereof) that are ejected after a significant 
time delay, we use the `delayed production' approximation (Pagel 1989) 
which assumes the element to be ejected at a fixed time delay $\Delta$ 
after the time of star formation, whereas the total mass ejection 
from a generation of stars is still assumed to take place instantaneously. 
In this case the abundance $z_{2}$ in units of the yield varies according to 
\eq
\frac{d}{dt}(gz_{2})=\omega'g'-(1+\eta)z_{2}\,\omega\,g; \;\; t>\Delta,  
\en
where the dashes indicate that the variable has to be taken at the 
time $t-\Delta$ rather than $t$.  Eq. (12) leads after some reduction to
\eq 
\frac{d}{dt}\left[z_{2}\left(e^{\eta u}-1\right)\right]  
=\omega' e^{(1+\eta)u} [e^{-u'} -e^{-(1+\eta)u'}].     
\en 

For smooth models ($\omega =$ const.), Eq. (13) has the solution 
\begin{eqnarray} 
z_{2}(u)&=&\frac{e^{(1+\eta)\,\omega\Delta}}{e^{\eta u}-1}
\left[\frac{e^{\eta (u-\omega\Delta)}-1}{\eta} - (u-\omega\Delta)
\right];\nonumber\\&&u>\omega\Delta. 
\end{eqnarray} 

\begin{figure} 
\epsfxsize=\hsize 
\epsfbox[85 85 510 297]{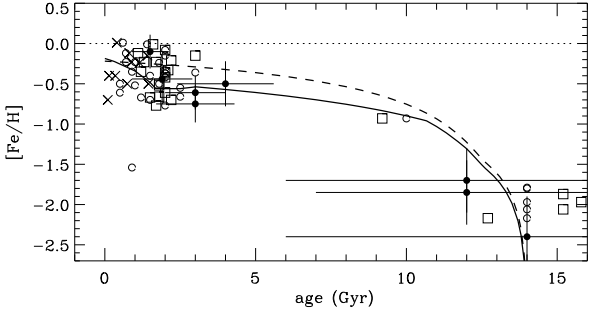} 
\caption{Age-metallicity relation for the LMC.  Curves represent 
our models as in previous figures.  Data sources are as follows: 
{\em open circles}, Olszewski {\em et al.} (1991); {\em crosses}, 
Girardi {\em et al.} (1995); {\em open squares}, Geisler {\em 
et al.} (1997); {\em filled circles}, Idiart \& 
de Freitas Pacheco (1997).} 
\label{fig4} 
\end{figure} 

\begin{figure}
\epsfxsize=\hsize 
\epsfbox[85 85 510 297]{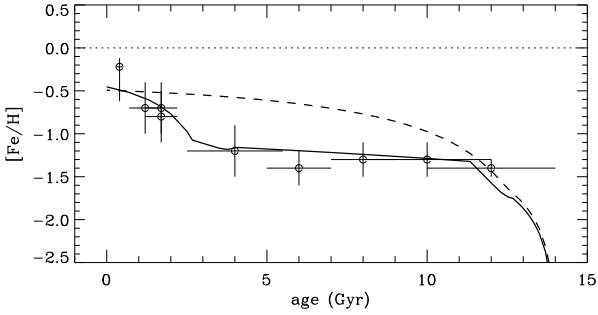} 
\caption{Age-metallicity relation for the SMC.  Curves represent 
our models as in previous figures.  Data represented by open 
circles are from Da Costa (1991). } 
\label{fig5} 
\end{figure} 

In bursting models, we assume $\omega = \omega_{1}$, say, up to time 
$t_{1}$, $\omega = \omega_{2}$ between $t_{1}$ and $t_{2}$ and 
$\omega =\omega_{3}$ between $t_{2}$ and $t_{3}=T$, the age of the
system, assumed to be 14 Gyr. $\omega_{2}$ is small, representing
a quiescent phase  between an initial starburst represented by 
$\omega_{1}$ and a stronger recent starburst represented by 
$\omega_{3}$. Then, in Eq. (13), 
\eq
\omega'=u' =0;\;\;{\rm for}\;t<\Delta;  
\en
\begin{eqnarray} 
\omega'&=&\omega_{1};\nonumber\\
u'&=&\omega_{1}(t-\Delta);\;\;{\rm for}\nonumber\\
\Delta &\leq & t\;<t_{1}+\Delta; 
\end{eqnarray} 
\begin{eqnarray} 
\omega'&=&\omega_{2};\nonumber\\
u'&=&\omega_{1}t_{1}+\omega_{2}(t-t_{1}-\Delta);\;\;{\rm for} 
\nonumber\\ 
t_{1}+\Delta &\leq & t\;<t_{2}+\Delta;
\end{eqnarray}
\begin{eqnarray}
\omega'&=&\omega_{3};\nonumber\\
u'&=&\omega_{1}t_{1}+\omega_{2}(t_{2}-t_{1})+\omega_{3} 
(t-t_{2}-\Delta);\;\;{\rm for}\nonumber\\  
t &\geq & t_{2}+\Delta. 
\end{eqnarray}
Also 
\begin{eqnarray}
u &=& \omega_{1}t;\;\;\;\;\;\;\;\;{\rm for}\; t<t_{1};\nonumber\\
&=& \omega_{1}t_{1}+\omega_{2}(t-t_{1});\;\;\;
{\rm for}\;t_{1}\leq t<t_{2};
\nonumber\\
&=&\omega_{1}t_{1}+\omega_{2}(t_{2}-t_{1})+\omega_{3}(t-t_{2});\;\; 
{\rm for}\;t\geq t_{2}. 
\end{eqnarray} 

Thus in separate time segments $\Delta$ to $t_{1}$ (or $t_{1}$ to 
$\Delta$ if $\Delta > t_{1}$), $t_{1}$ (or $\Delta$) 
to $t_{1}+\Delta$, $t_{1} 
+\Delta $ to $t_{2}$, $t_{2}$ to $t_{2}+\Delta $, $t_{2}+\Delta $ to 
$t_{3}=T $, $\omega' =$ const. and the exponents in Eq. (13) 
are all linear functions of time:
\begin{eqnarray} 
(1+\eta)u &=& p+qt; \nonumber\\ 
u' &=& r+st, 
\end{eqnarray} 
say, and so Eq. (13) can be solved analytically for each interval 
$t_{a}$ to $t_{b}$: 
\begin{eqnarray} 
&&\!\!\!\!\!\!\!\!\!\!\!\!\!z_{2}(t_{b})(e^{\eta u_{b}}-1) - z_{2}(t_{a}) 
\left( e^{\eta u_{a}}-1\right) \nonumber \\
&=&\omega'\frac{e^{p-r}}{q-s}\left\{e^{(q-s)t_{b}}
- e^{(q-s)t_{a}}\right\}\nonumber\\ 
&-&\omega'\frac{e^{p-(1+\eta)r}}{q-(1+\eta)s} 
\left\{e^{[q-(1+\eta)s]t_{b}}-e^{[q-(1+\eta)s]t_{a}}\right\}.   
\end{eqnarray}

\subsection{Yields and time delays} 

Table 3 gives the yields and time delays that we assume, after 
Pagel \& Tautvai\v{s}ien\.{e} (1995, 1997). Yields are in units of 
solar abundance of the corresponding element and time delays in Gyr 
are given in the top row.  Total abundances are obtained for each time
step by summing the products of these yields with the appropriate values
of $z$ for the relevant time delay, which are given in columns 7 to 11 
of Tables 1 and 2.  Figs 4 and 5 show the resulting 
age-metallicity relations for the two Clouds, which are similar 
to those derived by Tsujimoto {\em et al.} (1995), but differ somewhat 
from those for the LMC by Pilyugin (1996).

\begin{table} 
\centering 
\caption{Yields and time delays} 
\begin{tabular}{rccccc}
$\Delta$ (Gyr) &0&0.023&0.037&1.33&2.67\\ 
&\\ 
O&0.70&\\ 
Mg&0.88\\ 
Si, Ti&0.70&&&0.12\\ 
Ca&0.56&&&0.18\\
Fe&0.28&&&0.42 \\
&\\ 
Sr, La&0.013&0.11&0.29&&0.29\\ 
Y, Ba&0.010&0.08&0.30&&0.30\\ 
Zr&0.022&0.17&0.28&&0.28\\ 
Ce&0.020&0.16&0.26&&0.26\\ 
Pr, Sm&0.055&0.46&0.11&&0.11\\ 
Nd&0.040&0.33&0.17&&0.17\\ 
Eu, Dy&0.080&0.64&0.01&&0.01\\ 
\end{tabular} 
\end{table} 

\section{Element:element ratios} 
\begin{figure}
\epsfxsize=\hsize   
\epsfbox[40 40 566 765]{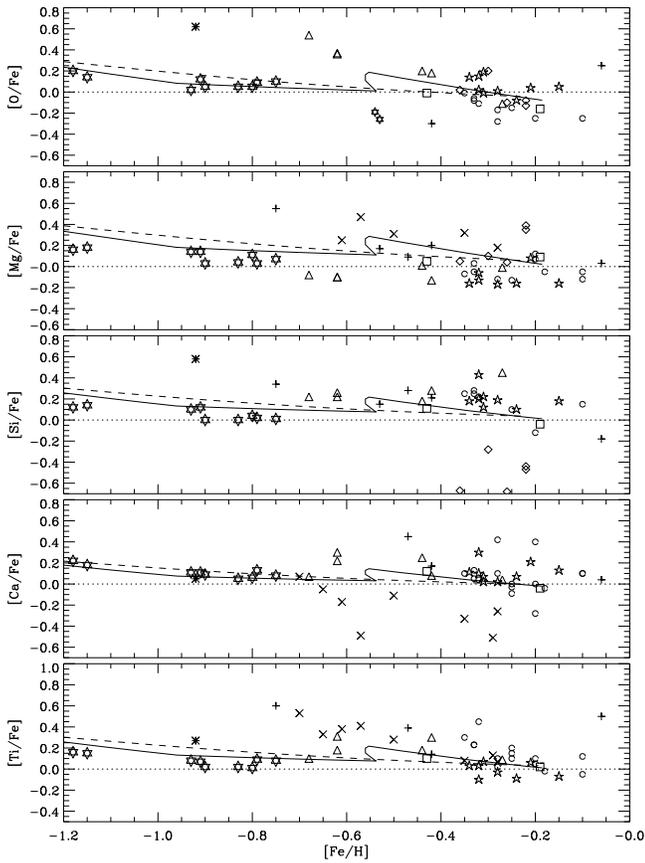}
\caption{Element:iron ratios for oxygen, $\alpha$-particle elements and 
Ti in the LMC and in the anomalous Galactic halo stars. The full-drawn 
curves show predictions from the bursting model and the broken-line 
curves those from the smooth model.  Data points 
for the LMC stars have been shifted upwards by 0.2 dex for oxygen 
and downwards by 0.1 dex for silicon.  Data sources: 
+ {\em  signs}, Luck \& Lambert (1992);
{\em open circles}, Th\' evenin (1997); {\em five-cornered stars}, Hill, 
Andriewski \& Spite (1995); {\em open squares}, Spite, Barbuy \& Spite 
(1993);
{\em open diamonds}, J\" uttner {\em et al.} (1992); {\em open triangles}, 
McWilliam 
\& Williams (1991); {\em small six-pointed stars}, Barbuy, de Freitas 
Pacheco \&
Castro (1994); {\em asterisks}, Richtler, Spite \& Spite (1989); 
{\em crosses}, Russell \& Bessell (1989); {\em large 
six-pointed stars}, anomalous Galactic halo stars after Nissen \& Schuster
(1997).}
\label{fig6} 
\end{figure} 
\begin{figure}
\epsfxsize=\hsize 
\epsfbox[40 40 566 765]{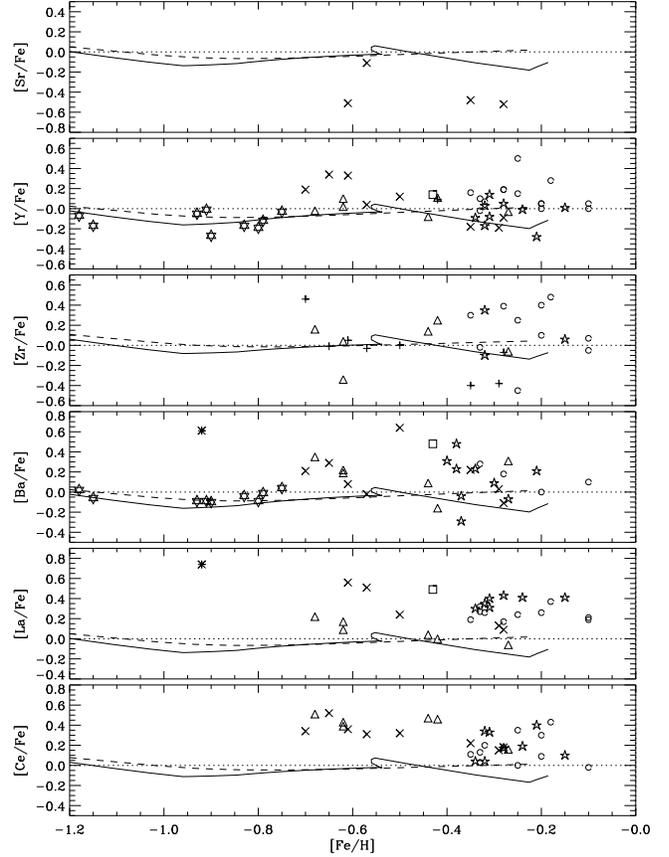} 
\caption{Element:iron ratios for s-process elements in the LMC and in 
the anomalous Galactic halo stars.  Curves and data 
sources as in Fig 6. Data points for Ba and La in the LMC stars have 
been shifted downwards by 0.1 dex and 0.16 dex respectively. } 
\label{fig7}
\end{figure} 

\begin{figure}
\epsfxsize=\hsize 
\epsfbox[40 40 566 765]{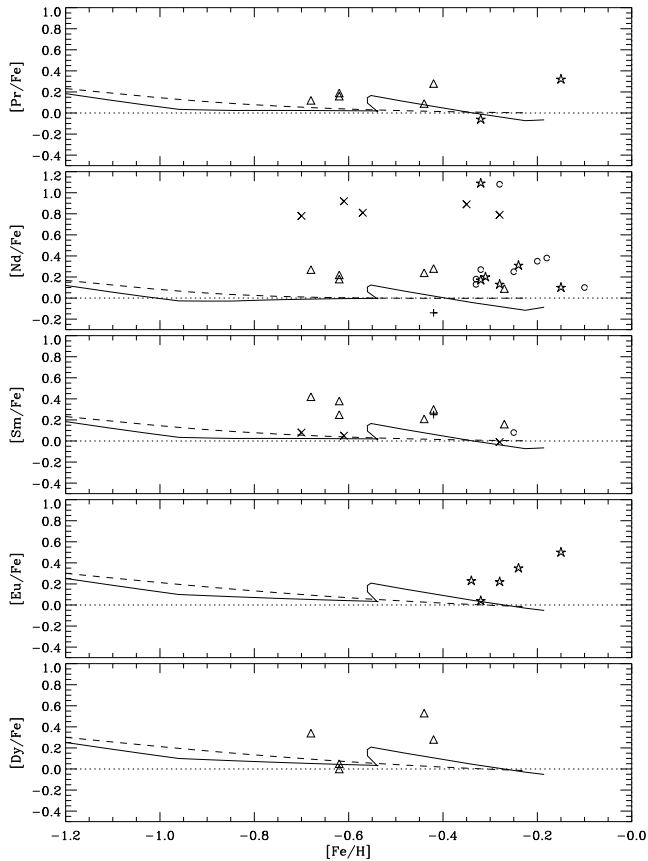} 
\caption{Element:iron ratios for heavy metals in the LMC.  Curves 
and data sources 
as in Fig 6.  Data points for Nd have been shifted downwards by 0.1 dex.} 
\label{fig8} 
\end{figure} 

\begin{figure}
\epsfxsize=\hsize 
\epsfbox[40 40 566 765]{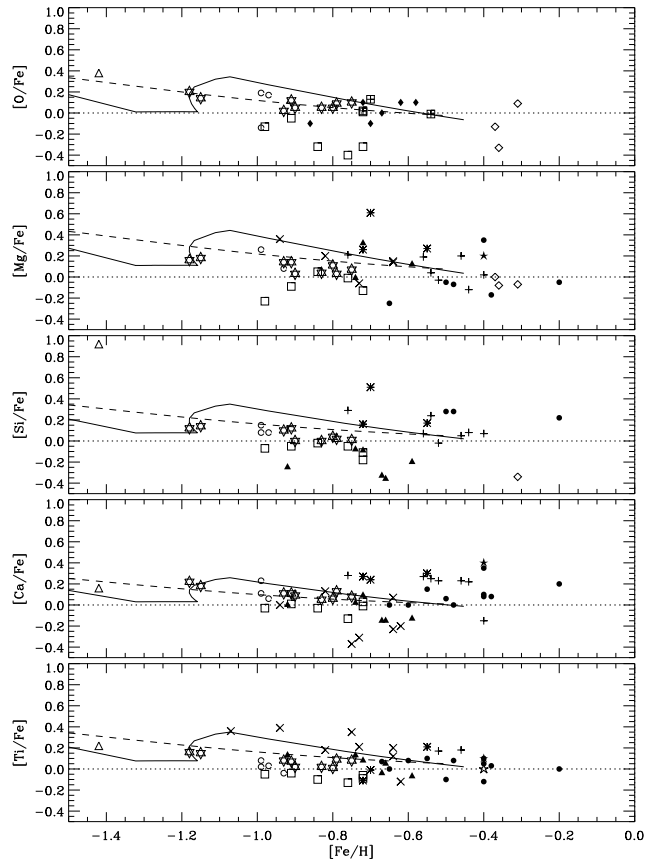} 
\caption{Element:iron ratios for oxygen, $\alpha$-particle elements and 
titanium for the SMC and anomalous Galactic halo stars.  Data points 
for oxygen and silicon in the SMC have been shifted upwards 
by respectively 0.2 dex and 0.1 dex .  Solid and broken curves represent 
our SMC bursting and smooth models respectively. Data sources: 
{\em open diamonds}, J\" uttner {\em et al.} (1992); {\em filled diamonds}, 
Hill, Barbuy
\& Spite (1997); {\em squares with} + {\em signs}, Spite, Barbuy \& Spite 
(1989); 
{\em open squares}, Hill \& Spite (1997); {\em open triangles}, 
Spite {\em et al.} 
(1986); {\em open circles}, Spite, Richtler \& Spite (1991); {\em filled 
circles},  
Th\' evenin (1997); + {\em  signs}, Luck \& Lambert (1992); {\em crosses}, 
Russell \& Bessell (1989); 
{\em asterisks}, Spite, Spite \& Fran\c cois (1989); {\em filled triangles}, Hill 
(1997); {\em filled five-pointed stars}, Th\' evenin \& Foy (1986); {\em 
open five-pointed stars}, Foy (1981); {\em open six-pointed stars}, 
anomalous Galactic halo stars from Nissen \& Schuster (1997).} 
\label{fig9}
\end{figure} 

\begin{figure} 
\epsfxsize=\hsize 
\epsfbox[65 65 566 595]{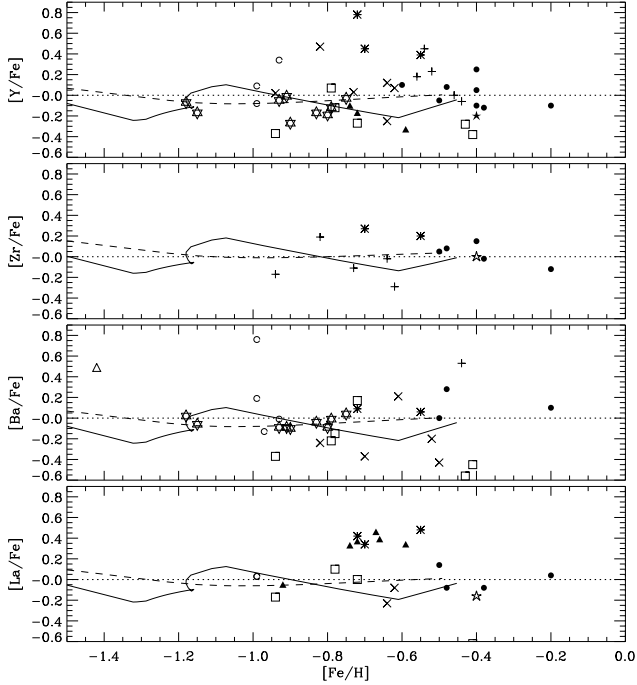} 
\caption{Element:iron ratios for s-process elements in the SMC and 
anomalous Galactic halo stars.  Curves and symbols as in Fig 9. Data 
points for Ba and La in the SMC stars have been shifted downwards by 
0.1 dex and 0.16 dex respectively.} 
\label{fig10}
\end{figure} 

\begin{figure}
\epsfxsize=\hsize 
\epsfbox[65 65 566 595]{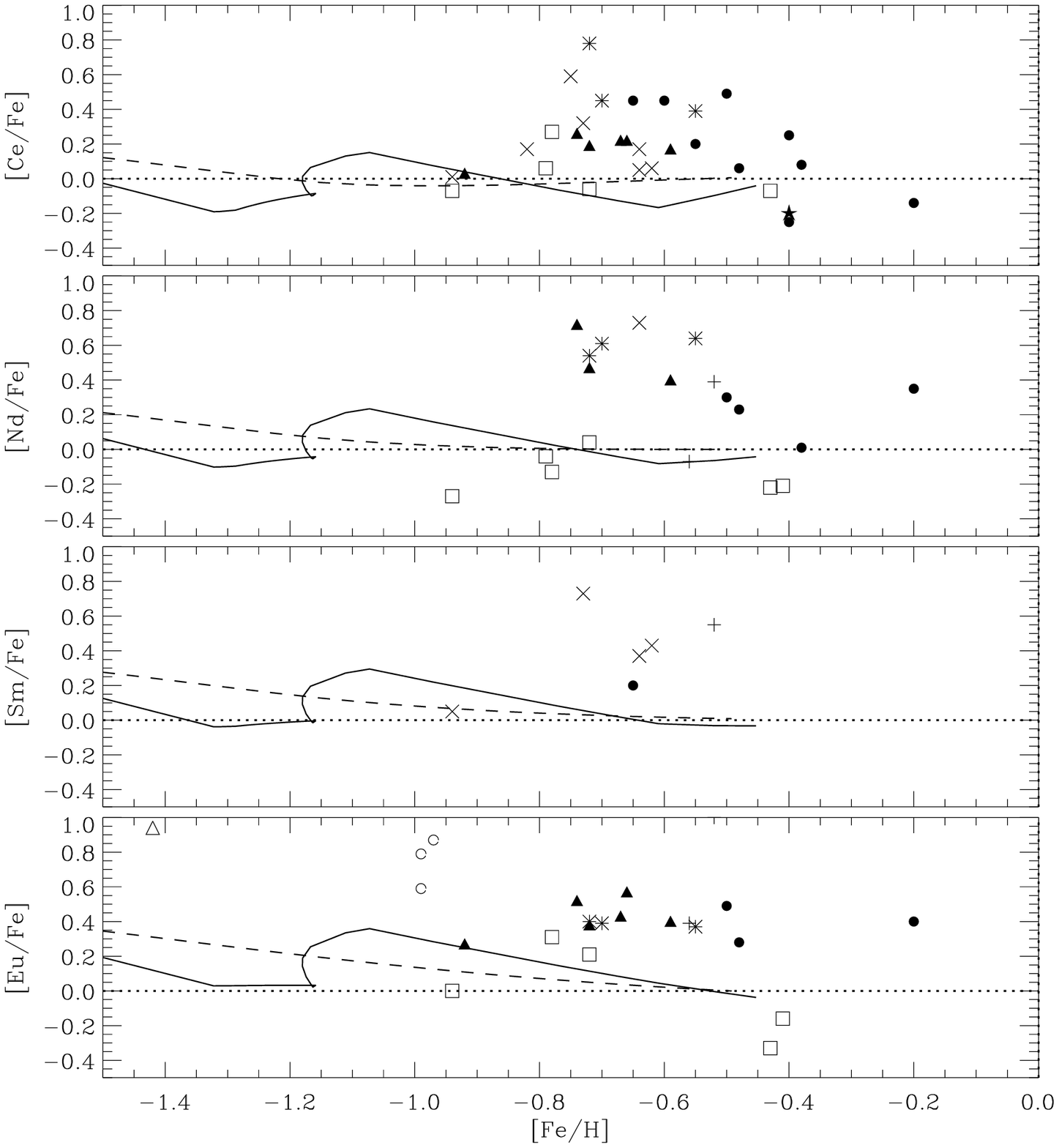} 
\caption{Element:iron ratios for heavy metals in the SMC.  Curves and
symbols as in Fig 9. Data points for Nd have been shifted downwards by 0.1
dex.} 
\label{fig11}
\end{figure}

The last seven columns of Tables 1 and 2 give abundances or 
constituent $z$-values for individual elements in the 
Clouds calculated from our bursting models, and element:iron ratios are 
plotted in Figs 6 to 11.  The observational data, which come mainly 
from MC supergiants, have in some cases 
been adjusted to be relative to Galactic supergiants rather than the 
Sun, since (as noted by Russell \& Dopita 1992, Pagel 1992 and 
Hill, Andrievsky \& Spite 1995) 
there are some discrepancies, particularly in the case of oxygen, and 
it seems better to compare like with like. Russell \& Dopita took 
[O/Fe] for the solar neighbourhood to be $- 0.03$ only, but various studies 
of Galactic supergiants (Luck \& Lambert 1985; Spite, Barbuy \& Spite 
1989; Hill, Andrievsky \& Spite 1995; Venn 1995) indicate a more 
negative value, [O/Fe] $\simeq -0.2$ relative to the Sun. The observational 
[O/Fe] ratios plotted in Figs 6 and 9 have accordingly been shifted 
upwards by 0.2 dex, which leads to a somewhat different picture from 
that assumed in previous chemical evolution models for the Clouds; in 
particular, there is now (apparently) good agreement between oxygen 
and the other $\alpha$-elements.  The oxygen abundances from planetary 
nebulae shown in Fig 2 also support this upward adjustment. 
On the basis of studies of Canopus 
and other Galactic supergiants by Luck (1982), Reynolds, Hearnshaw 
\& Cottrell (1988), Russell \& Bessell (1989) and Spite, Spite \& 
Fran\c{c}ois (1989), we have shifted the data points for [Si/Fe], 
[Ba/Fe] and [Nd/Fe] downwards by 0.1 dex and [La/Fe] downwards by 
0.16 dex, taking all the remaining elements at face value.  

Although the model curves in Figs 6 to 11 are meant to represent an 
evolution in time, it has to be borne in mind that the data points 
come from young supergiants, so that their horizontal spread is due 
to scatter (part of which is probably observational) and not from 
evolution in time.  Consequently it is only the centroid of these 
points that is significant and no evolutionary trend can be deduced 
from the data.  However, 
Nissen \& Schuster (1997) have found a sample of stars from the outer 
halo of our Galaxy in which, among other peculiarities, the 
$\alpha$-element to iron ratios are more or less solar rather than 
enhanced as in other stars with the same metal deficiency.  They 
suggest that these stars could have been captured from dwarf galaxies 
such as the Magellanic Clouds and we accordingly include them (without 
any adjustments) in the relevant plots.  Our LMC model fits these 
ratios very well, including a slight rise from solar-like [$\alpha$/Fe]
around [Fe/H] $=-0.8$ to [$\alpha$/Fe] $\simeq 0.2$ at 
[Fe/H] = $-1.2$ (Fig 6) 
and a more uniform trend for s-process elements (Fig 7), supporting an 
origin in the LMC (or some similar system).  The fit of our SMC 
model (Figs 9, 10) is not quite as good.  
As far as the actual Magellanic stars are concerned, there is considerable 
scatter in the data, much of which we suspect to be unreal, although we 
have chosen to plot individual determinations rather than to follow 
Russell \& Dopita in just plotting an average.  Within 
the scatter, the agreement of our model with abundances deduced from 
observation seems to be quite satis\-factory, with the possible 
exception of a few of the heaviest elements.  Broadly speaking, there 
is hardly any evidence for significant departures from solar, or 
solar-neighbourhood, abundance ratios among the elements considered
(we do not discuss carbon, nitrogen or sodium which all exhibit 
`secondary' behaviour), and that result is just what is predicted;   
whereas the steepened IMF models of Tsujimoto {\em et al.}, while  
successful for oxygen (if unadjusted to Galactic supergiants), 
predict subsolar abundances (relative to iron) of several 
$\alpha$-particle elements, notably magnesium, which are not supported 
by the data.  Pending improved data, especially for the heaviest 
elements, we conclude that neither a steepened IMF nor selective 
galactic winds are required to explain the abundances in the Magellanic 
Clouds.  

\section{Relative supernova rates} 

Barbuy, de Freitas Pacheco \& Castro (1994) and Tsujimoto {\em et al.} 
(1995) have claimed that the overall ratios of Type Ia supernovae to 
core-collapse supernovae ever formed in the Clouds are higher than in 
the solar neighbourhood.  Our models do not support this and they 
fall down if it is true.  With our approximations, the ratio is 
proportional to 
\eq
\frac{s(T-\Delta)}{s(T)}=\frac{m(T-\Delta)-g(T-\Delta)}
{m(T)-g(T)}. 
\en
Taking $u$ = 4.5, $\omega\Delta = 0.4$ for the solar neighbourhood 
(Pagel \& Tautvai\v{s}ien\.{e} 1995), this ratio comes out to be 0.95, 
whereas in the LMC and SMC according to Tables 1 and 2 it is 0.86 
and 0.77 respectively, i.e. marginally lower. The situation changes if 
we compare the above Magellanic ratios with those that prevailed in 
the solar neighbourhood at those times in the past when it had the
corresponding metallicities; these turn out to be 0.81 for [Fe/H] 
$=-0.2$ and 0.62 for [Fe/H] $=-0.45$.   Thus we attribute the 
lower $\alpha$/Fe ratios chiefly to the effect of slower evolution up to 
a metallicity which prevailed in our Galaxy at an earlier time 
when there had indeed been relatively fewer SNIa in the latter.  
It can be seen from Figs 6 to 11 that the influence 
of starbursts on the element:element ratios is comparatively 
insignificant in our models. 

When we consider {\em current} supernova rates, the situation is 
quite different again.  In this case the ratio is proportional to 
\eq
\frac{(ds/dt)_{T-\Delta}}{(ds/dt)_{T}}=\frac{g(T-\Delta)}{g(T)}. 
\en
The corresponding numbers are 1.2 for the solar neighbourhood, 1.8 
for the LMC and 1.5 for the SMC in our bursting models.  Thus we 
expect a 50 per cent higher ratio of SNIa to core-collapse supernovae 
in the LMC at present.  Assuming a ratio in the solar neighbourhood 
between 0.11 and 0.25 (van den Bergh \& McClure 1994), the ratio in the 
LMC is then predicted to be between 0.16 and 0.38, which is in good 
agreement with the average figure of 0.29 given for Sdm-Im galaxies 
by Cappellaro {\em et al.} (1993).  A still higher ratio,
of order 1, has been estimated for the LMC on the basis of  
X-ray observations of remnants by Hughes {\em et al.} (1995), 
who suggest a likely lower limit of 0.25, so that our model agrees 
qualitatively, if not quantitatively, with their findings. 
Current ratios are sensitive to the precise assumptions made 
about the bursts, and do not necessarily represent historic averages.
We suggest that the relatively high ratio of SNIa to core collapse 
SN observed in the LMC and other Sd-Im galaxies is related  
to their star formation history rather than the IMF. 

\section*{Acknowledgments}

GT acknowledges a NORDITA Baltic/NW Russia Fellowship from the 
Nordic Council of Ministers and hospitality of NORDITA while 
the work for this paper was being carried out.

\bsp
\end{document}